\documentclass[11pt,twocolumn]{article} 
\usepackage{graphicx}  
\usepackage{amsmath}
\usepackage{bm}        
\usepackage{amssymb}   
\usepackage[margin=0.75in]{geometry}
\usepackage{caption} 
\usepackage{float}
 
\begin{document}

\title{Highly cooperative stress relaxation in two-dimensional soft colloidal crystals}
\date{\today}
\author{B. van der Meer$^{1,2}$, W. Qi$^{2}$, R. Fokkink$^{1}$,\\ J. van der Gucht$^{1}$, M. Dijkstra$^{2}$ and J. Sprakel$^{1}$\\
\\$^1$Laboratory of Physical Chemistry and Colloid Science,\\
Wageningen University, Dreijenplein 6, 6703 HB Wageningen, The Netherlands
\\$^2$Soft Condensed Matter, Debye Institute for Nanomaterials Science,\\ 
Utrecht University, Princetonplein 5, 3584 CC Utrecht, The Netherlands\\}

\maketitle
\begin{abstract}
\textbf{Stress relaxation in crystalline solids is mediated by the formation and diffusion of defects. While it is well established how externally-generated stresses relax, through the proliferation and motion of dislocations in the lattice, it remains relatively unknown how crystals cope with internal stresses. We investigate, both experimentally and in simulations, how highly localized stresses relax in two-dimensional soft colloidal crystals. When a single particle is actively excited, by means of optical tweezing, a rich variety of highly collective stress relaxation mechanisms results. These manifest in the form of open strings of cooperatively moving particles through the motion of dissociated vacancy-interstitial pairs, and closed loops of mobile particles, which either result from cooperative rotations in transiently generated circular grain boundaries or through the closure of an open string by annihilation of a vacancy-interstitial pair. Surprisingly, we find that the same collective events occur in crystals which are excited by thermal fluctuations alone; a large thermal agitation inside the crystal lattice can trigger the irreversible displacements of hundreds of particles. Our results illustrate how local stresses can induce large scale cooperative dynamics in two-dimensional soft colloidal crystals and shed new light on the stabilisation mechanisms in ultrasoft crystals.}
\end{abstract}
\newpage
\clearpage

\section{Introduction}
Stress relaxation in crystalline solids is governed by the formation and diffusion of defects in the crystal lattice. For small deformations, it is well known that relaxation occurs through the motion of sparse dislocations~\cite{taylor1934mechanism,polanyi1934art,orowan1934kristallplastizitat,schall2004visualization,schall2006visualizing}. However, it remains unclear how a crystalline solid copes with stresses that are generated well-inside the crystal; either caused by external sources ~\cite{bai2010efficient,siders1999detection} or by thermal excitations, which can become especially important in superheated states~\cite{jin2001melting,wang2012imaging,bai2008ring,gallington2010thermodynamic,zhang2013string}. Particle rearrangements that result from large internal perturbations must necessarily involve the motion of many of the constituent particles simultaneously. Often these collective dynamics are rare due to large activation barriers in the dense solid state. As a result, studying large-scale collective dynamics inside crystalline solids is challenging. One may expect that sufficiently large fluctuations, which could drive collective rearrangements, may only appear when the elastic energy associated with a fluctuation becomes on the order of the thermal energy. In crystals formed from colloidal particles that interact through long-range repulsive interactions, low-density and ultrasoft solid states are experimentally accessible in which large thermal excitations can be easily observed using optical microscopy~\cite{yethiraj2003colloidal}. While it remains unclear how such fragile solids respond to large agitations inside the crystal lattice, identifying the microscopic mechanisms of stress relaxation in relation to the mechanical stability of solids is of fundamental importance to understand phenomena like creep, yield and fracture.

These colloidal model systems allow manipulation of the kinetic states of individual particles by means of optical tweezers, for example to create vacancies and interstitials ~\cite{pertsinidis2001equilibrium,pertsinidis2001diffusion,kim2011optical, libal2007point,dasilva2011mechanism}, or to manipulate many-particle defect reactions ~\cite{irvine2013dislocation}. However, the response of colloidal crystals to large thermal and external excitations of a single particle within the lattice is largely unexplored. As a result, the relationship between stress relaxation mechanisms, in response to internal perturbations, and the ultimate stability of the solid phase remains poorly understood.  

In this paper, we investigate how stresses relax in two-dimensional soft colloidal crystals using a combination of experiments and computer simulations. When a single particle inside the crystal is actively driven out-of-equilibrium, a rich variety of collective stress relaxation mechanisms result, mainly in the form of open and closed strings of rearranging particles. Surprisingly we find that these unusual collective rearrangements are not restricted to crystals that are actively perturbed, but also appear in soft colloidal crystals excited through thermal fluctuations alone. A sufficiently large internal agitation inside the lattice can cause the irreversible rearrangement of hundreds of particles from their previous equilibrium positions. These results illustrate the complexity of internal stress relaxation through collective and activated modes, and shed new light on the origins of stability and instability in marginally stable crystalline solids.

\section{Results}
We study stress relaxation in two-dimensional colloidal crystals, formed spontaneously by charged colloidal particles that are confined to two dimensions by gravity. The particles with diameter $\sigma =$ 2.8 $\mu m$ interact through long-ranged, Yukawa-like, electrostatic repulsions, with a screening length $\kappa=0.8$ $\mu m^{-1}$, and form a hexagonal crystal phase at area fractions as low as $\phi_A \approx 0.13$. In these low-density crystals, lattice spacings can be as large as $a \approx$ 6-8 $\mu m$. Using optical tweezers, a single particle in these low-density crystals can be trapped and manipulated with high fidelity. Additionally, we perform Brownian Dynamics (BD) simulations of two-dimensional crystals of $N=2500$ Yukawa particles in which a particle is either actively driven by an external force or by thermal fluctuations alone. We parameterise the potential using the experimental data, measured as described in~\cite{masri2011measuring}, and find good agreement with previously reported values for these systems~\cite{hsu2005charge,roberts2008electrostatic}, yielding a contact value of $\beta U_0=235$ for the Yukawa potential with $\beta =1/k_BT$ and $\kappa \sigma =2.25$. 

In both our experiments and simulations, the response of these two-dimensional crystals upon a sinusoidal oscillation of a single particle depends strongly on the amplitude $A$ of the perturbation. In the following, we express the amplitude of the perturbation normalised to the lattice constant $a$: $\gamma_L = \langle|{\bf u}_{tracer}|^2 \rangle^{1/2}/a \approx 2A / \pi a $, where $\langle|{\bf u}_{tracer}|^2 \rangle^{1/2}$ is the root-mean square displacement of the tracer from its equilibrium position in the lattice. At low amplitudes, $\gamma_L < 0.5$, the crystal responds elastically; the driven particle pushes the particles in front of it along an one-dimensional path, in which no irreversible particle rearrangements occur. At higher amplitudes a transition from elastic to plastic deformation is observed, manifested by a region dense in irreversible particle rearrangements surrounding the driven particle [Figure S1 $\&$ S2]. In this region, the particles exhibit no crystalline order and exhibit a high mobility, which is reminiscent of local, mechanically-induced, melting~\cite{dullens2011shear,reichhardt2004local}.

\begin{figure}
\begin{center}
\includegraphics[width=0.5\textwidth]{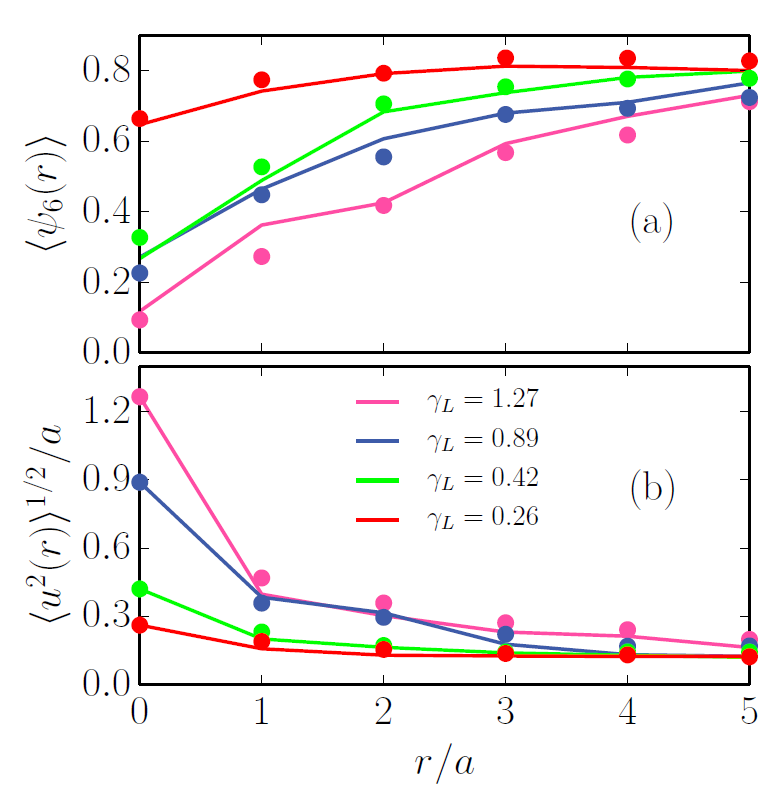}
\caption{Bond-orientational order $\langle \psi_6 (r) \rangle$ (a) and root-mean square displacement $\langle |{\bf u}(r)|^2 \rangle^{1/2}/a$ (b) as a function of the normalized distance $r/a$ to the excitation, for four different normalized perturbation amplitudes $\gamma_L$. Symbols represent experimental data and drawn lines the results from BD simulations. All data points were binned to distances of integer lattice spacings to obtain sufficient statistics for averaging.\label{fig:bopandmsd}}
\end{center}
\end{figure}

The length scale over which the crystalline order is lost and particle mobility is increased depends strongly on the normalized perturbation amplitude $\gamma_L$, both in experiments and in simulations. We first explore the loss of structure using the local bond-orientational order parameter of particle $i$, which is given by $ \psi_{6i} = \sum_{j=1}^{\mathcal{N}_i} e^{6i\theta_{ij}} / \mathcal{N}_i $, where $\theta_{ij}$ is the angle of the bond between particles $i$ and $j$ relative to an arbitrary reference axis~\cite{steinhardt1983bond}. We plot the average local bond orientational order parameter $\langle \psi_{6}(r) \rangle$ as a function of the distance $r$ to the probe as obtained from the experiments (symbols) as well as the simulations (lines) in Figure \ref{fig:bopandmsd}(a). For small sinusoidal perturbations, i.e., $\gamma_L<0.5$, the structural damage to the crystal lattice is minimal, reaching only the nearest neighbours of the driven colloid, while at normalized perturbation amplitudes $\gamma_L>0.5$ the positional order is lost over an area spanning several lattice constants. Brownian dynamics simulations yield results that are in quantitative agreement with the experimental data; thus ruling out hydrodynamic or laser-induced effects as the cause of the local lattice disruption. The transition from elastic to plastic deformation is also reflected in the root-mean square particle displacement. Only for higher normalized perturbation amplitudes $\gamma_L>0.5$ the particles exhibit increased mobility, through irreversible particle rearrangements [Figure \ref{fig:bopandmsd}(b)]. The activated dynamics leading to enhanced mobility and the loss of crystallinity are evidently related; the large agitations that we induce require many particles to displace from their lattice sites, destroying the local positional order.

To disentangle the formation of a locally mobile zone from perturbation-induced vitrification, in which the lattice order is lost and the dynamics is kinetically arrested, we use the two-dimensional equivalent of the Lindemann criterion, which is usually used in the context of 2D melting~\cite{zahn1999two,han2008melting,qi2010melting}. The, empirical, Lindemann criterion states that a crystal becomes unstable due to vibrations when the amplitude of positional fluctuations of a particle around its mean position exceeds a certain fraction of the lattice spacing $a$~\cite{lindemann1910ueber}.  For two-dimensional crystals, however, the mean square displacement $\langle |{\bf u}|^2 \rangle$ diverges due to strong long-wavelength fluctuations~\cite{mermin1968crystalline}. Therefore we use the modified definition of the Lindemann parameter $L_{m,i}=\langle |{\bf u}_i-{\bf u}_j|^2 \rangle^{1/2}/{a}$, where ${\bf u}_i-{\bf u}_j$ is the relative displacement of neighbouring particles $i$ and $j$~\cite{bedanov1985modified}. 

\begin{figure}
\begin{center}
\includegraphics[width=0.5\textwidth]{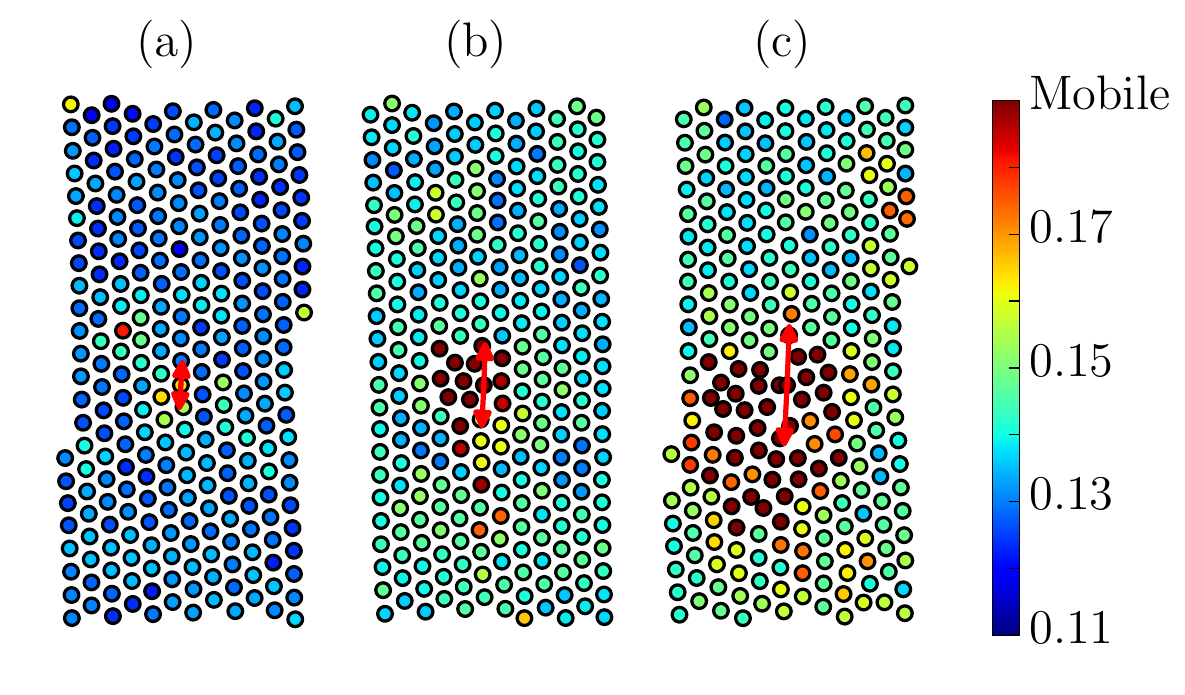}
\caption{\label{fig:fig2} Reconstructed particle configurations, in which the particles are colored according to their Lindemann parameter $L_m$, for three different strain amplitudes: $A = 2.4$ $\mu m $ ($\gamma_L =0.25$) (a), $A = 7.2$ $\mu m$ ($\gamma_L =0.8$) (b) and $A = 12.0$ $\mu m $ ($\gamma_L =1.2$) (c).}
\end{center}
\end{figure}

\begin{figure}
\begin{center}
\includegraphics[width=0.5\textwidth]{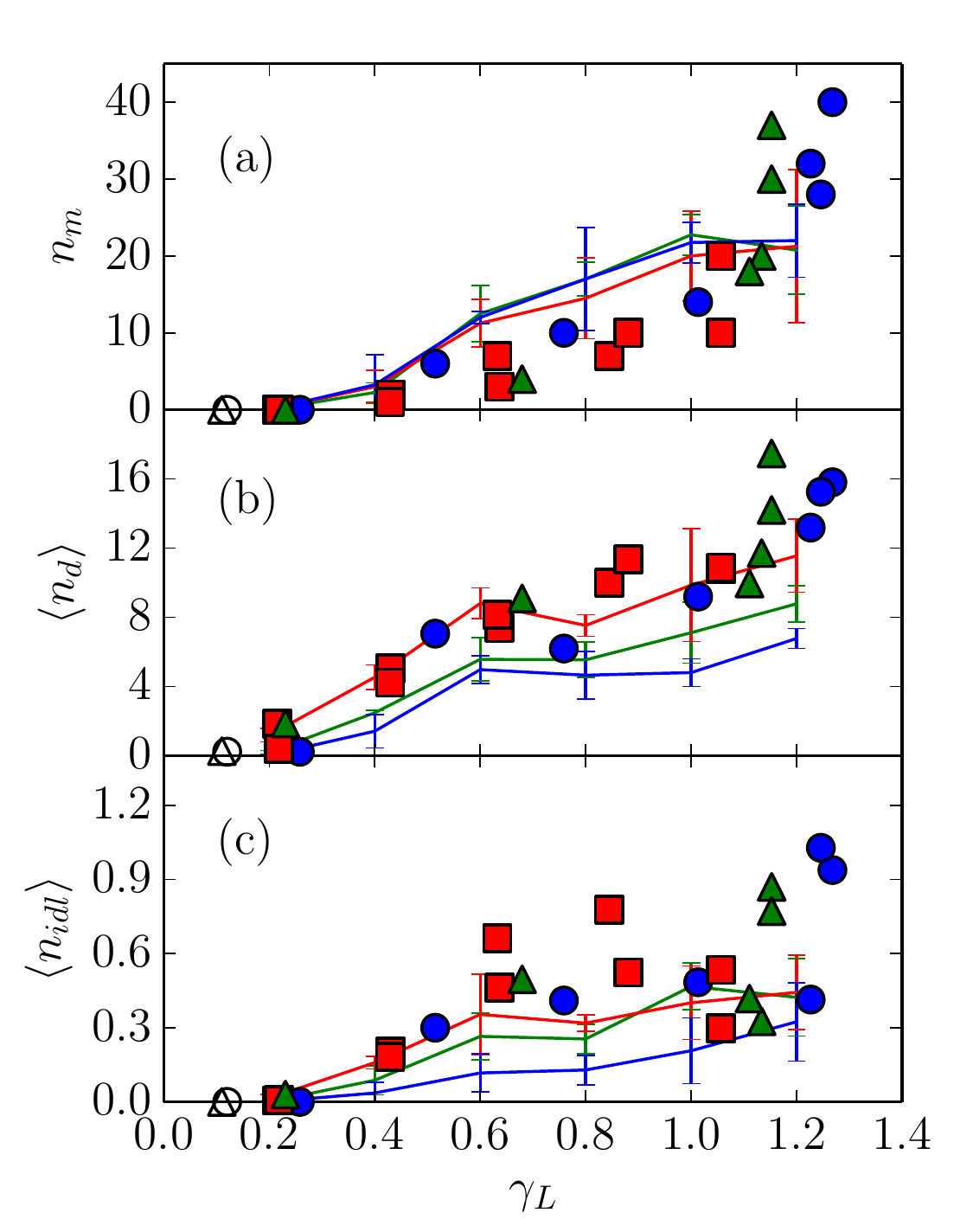} 
\caption{\label{fig:fig3} Number of mobile particles $n_m$ (a), average number of defects $\langle n_d \rangle$ (b) and average number of isolated dislocations $\langle n_{idl} \rangle$ (c) as a function of reduced strain $\gamma_L$, for area fractions $\phi_A=$ 0.19 (circles), 0.16 (triangles) and 0.14 (squares). Symbols are experimental data, open symbols represent data for non-perturbed, purely thermal, crystals and drawn lines are the results from BD simulations.} 
\end{center}
\end{figure}

For $\gamma_L<0.5$, the vibrations within the crystal remain fairly homogeneous, as shown by the reconstructed and color-coded Lindemann maps in Figure \ref{fig:fig2}(a) $\&$ S3(a) for experiments and simulations, respectively. However, for $\gamma_L>0.5$, a mobile and disordered zone forms which grows with increasing normalized perturbation amplitude $\gamma_L$ [Figure \ref{fig:fig2}(b,c) $\&$ S3(b,c)]. To further quantify this observation, we choose a critical Lindemann value of $L_m^*=0.19$ to distinguish between crystalline and mobile particles. From our experimental data, we observe that no crystals exists with $L_m>0.19$, hence we use this as the criterion to identify the mobile particles. We note that all trends are robust to variations in the choice of $L_m^*$. We find that the size of the mobile zone, expressed as the number of mobile particles $n_m$ in this region, is zero at low normalized perturbation amplitudes; beyond a critical amplitude of $\gamma_L^* \approx 0.3$ a mobile region first appears which subsequently grows with increasing $\gamma_L$ [Figure \ref{fig:fig3}(a)]. Surprisingly, normalisation of the perturbation amplitude with the lattice constant leads to a collapse of the curves for three different concentrations. Clearly, both the onset and spatial extent of the mobile zone is governed primarily by the ratio of local perturbation amplitude to the lattice spacing.

As the crystal becomes unstable, the lattice is disrupted and defects, identified as particles for which the number of nearest-neighbours is unequal to six, start to proliferate. This is shown by a strong increase in the average number of defects, $\langle n_d \rangle$, in Figure \ref{fig:fig3}(b). The breaking of the local symmetry and the onset of mobility are evidently related. At sufficiently high driving amplitudes $\gamma_L$  we also observe a large increase in the average number of energy-costly isolated dislocations $\langle n_{idl} \rangle$ [Figure \ref{fig:fig3}(c)].    

The question remains what dynamics on the scale of individual particles mediate the formation of costly defects and ultimately lead to extraordinary particle mobility. Time-lapse sequences of both experimental and simulated crystals, when perturbed by applying an external oscillatory force to  a randomly selected particle with a sufficiently high amplitude $\gamma_L$, show the emergence of cooperative dynamics in the form of string-like rearrangements in which particles take over the positions of their neighbors in a sequential manner. These rearrangement chains may grow from both ends; strings grow both at the head, where particles are compressed ahead of the driven tracer, as well as from the tail, where particles start to explore the free space that was previously occupied. In some cases, the ends of such a string of cooperatively moving particles meet, resulting in a closed loop of rearrangements.

\begin{figure}
\begin{center}
\includegraphics[width=0.5\textwidth]{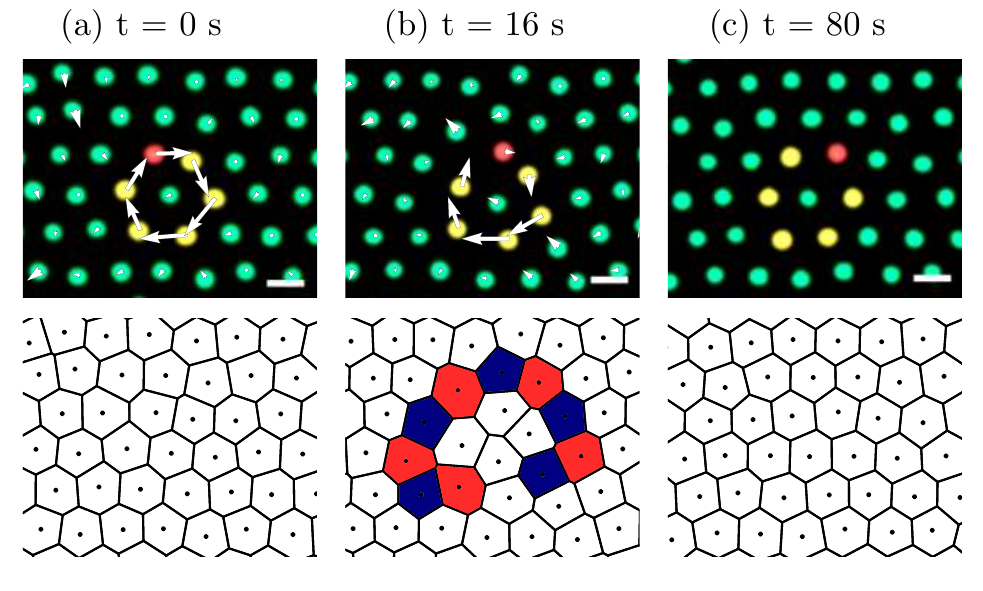}
\caption{\label{fig:fig4} Time lapse sequence of color-coded bright-field images (top) of a collective rearrangement loop; showing the driven tracer (red), the particle participating in the collective motion (yellow) and those that remain unperturbed (green). Vectors indicate the residual displacement between $t$ and the time at which the cooperative motion completes. Bottom row shows corresponding Voronoi tessellations in which the defects are highlighted using blue and red cells for 5-fold and 7-fold coordinated particles, respectively. Perturbation: $\gamma_L =0.8$. Scale bar: 5.0 $\mu m$.}
\end{center}
\end{figure}

One example of a particle rearrangement loop is shown in Figure \ref{fig:fig4} (movie shown in SI). The displacement vectors, superimposed on color-coded images (top row), indicate the residual displacement from the current time (indicated above the images) to the time when the sequence is completed at t = 80s. When the tracer particle, shown in red, is driven away from its equilibrium position, a loop of rearrangements of particles occurs, which relieves the stresses that are generated. The particles that are involved in this collective motion, shown in yellow, sequentially take over the positions of their predecessors in the loop. The chain of collectively moving particles grows from both the tail and head. Once the loop has closed, the crystal is temporarily restored, until further excitations can initiate new string-like rearrangements. 

We observe these anomalous cooperative modes both in our experiments and computer simulations; they emerge in a wide variety of different configurations, as illustrated in Figure S4 $\&$ S5. The loops we find typically include between 3 to 16 particles. Sometimes these loops encircle other, stationary, particles. After closure of a loop, the particles recrystallize, without any memory effects. The defect pattern associated with these relatively small loop-like rearrangements often follows a similar progression irrespective of the exact shape of the loop. In the Voronoi tesselations in the bottom row of Figure \ref{fig:fig4} particles with 5 or 7 nearest-neighbors are represented as blue and red cells, respectively. During a closed loop of particle rearrangements several defects are generated, which organise in a single string of multiple 5-7 pairs, forming a circular grain boundary [Figure \ref{fig:fig4}(b)]. The highly localized and cooperative particle motions form an efficient pathway to annihilate these defects and re-establish the crystalline structure [Figure \ref{fig:fig4}(c)].  

In addition to the closed loops of rearranging particles, we often observe open-ended strings of collective motion that do not close within the time frame of our experiment. Even when the trapped particle is returned to its equilibrium position, the string of cooperative motion that has been initiated may persist. In Figure S4(b), the red particles form such a chain in which the head and tail of this dynamic structure are in continuous search for new equilibrium positions. 

\begin{figure*}
\begin{center}
\includegraphics[width=0.88\textwidth]{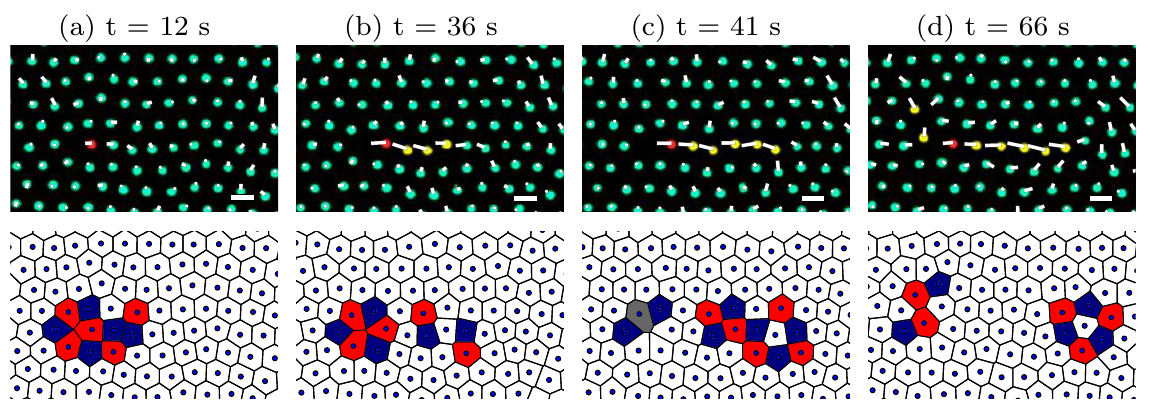}
\caption{ Time lapse sequence of color-coded bright-field images (top) and corresponding Voronoi tesselations (bottom) of an open-ended rearrangement string, which grows from both ends. Color coding as in Figure \ref{fig:fig4}. We have superimposed lines that connect the current particle positions with the initial defect-free configuration at $t = 0 s$. Bottom row shows corresponding Voronoi tessellations in which the defects are highlighted using blue, red and grey cells for 5-fold, 7-fold and 8-fold coordinated particles, respectively.  Perturbation: $\gamma_L =0.5$. Scale bar: 5.0 $\mu m$. \label{fig:stringgrowthexp}}
\end{center}
\end{figure*}

A time-lapse sequence of how such an open-ended chain of collectively rearranging particles grows is shown in Figure \ref{fig:stringgrowthexp} $\&$ S6 for our experiments and simulations, respectively. The superimposed displacement lines indicate the displacement from the initial, defect-free configuration at t=0s. When the trapped particle, shown in red, is driven sufficiently far away from its equilibrium position, the surrounding particles start to exhibit cooperative motion (yellow). Particles at the head of the string hop to nearby cages, thus joining the cooperative motion and making the string grow, whereas in a later stage the particles at the tail of the string also start to break from their own cages and explore the vacant site.

Also for the open-ended string-like rearrangements, the defect pattern appears robust to changes in the size and conformation of the rearrangement string [bottom rows of Figure \ref{fig:stringgrowthexp} $\&$ S6]. Once the particle is dragged sufficiently far away from its original lattice site a bound vacancy-interstitial pair first appears [Figure \ref{fig:stringgrowthexp}(a) $\&$ S6(a)]. Both vacancies and interstitials have many distinct topological configurations with different symmetries~\cite{pertsinidis2001equilibrium,libal2007point}. In the first snapshot the vacancy, consisting of three 7-5 defects, has three-fold symmetry and the interstitial, containing two 7-5 defects, has two-fold symmetry. When the particle is dragged even further from its original lattice site, the vacancy-interstitial pair unbinds and the resulting point defects start to migrate through the crystal [Figure \ref{fig:stringgrowthexp}(b-d) $\&$ S6(b-d)]. The interstitial is always located at the head of the string, whereas the vacancy is present at its tail. The movements of these point defects form the microscopic mechanism for the growth of the string, allowing strings to grow from both the interstitial and the vacancy-end. As predicted in previous computer simulations, we also observe that the migration of these point defects is accompanied by switching between different topological defect configurations~\cite{libal2007point,dasilva2011mechanism}. Our simulations show that these open-ended rearrangement strings have a profound influence on the single particle mobility; long simulations reveal that a single open-ended string caused by a highly localised perturbation can lead to appreciable diffusion in which many particles ($N \gg 500$) exchange irreversibly lattice sites.

In the majority of our simulations, however, the vacancy-interstitial pair recombines relatively fast after dissociation, giving rise to a second type of closed loop rearrangement, mediated by dissociation-recombination reactions of vacancies and interstitials rather than through the formation of circular grain boundaries. After unbinding, both the vacancy and interstitial migrate through the crystal for a while, but remain in each others proximity, and subsequently recombine quickly leaving a perfect lattice behind with no residual topological distortions. The closure of such an open-ended string is driven by the short-range attraction between a vacancy and an interstitial~\cite{kim2011optical}. Even though these loops can have the same size and shape, it appears that they proceed through a different evolution of the topological structure of the solid. Alternatively, the absence of a clear vacancy and interstitial in the grain-boundary mediated rearrangements loops might be due to an ``overlap" between the point defect topologies. We note, however, that the loops that proceed through circular grain boundaries seem to move in a more joint-like fashion compared to the loops that proceed through dissociated Frenkel pairs, for which the rearrangement string grows in a more sequential fashion at the head and tail only.   

In some cases, the growth of open-ended rearrangement strings follows a more complex process [Figure S7 and movie in SI]. The displacement lines, superimposed on color-coded images, indicate the displacement from the initial configuration at t = 0s. The tracer particle pushes the particles in front of it from their equilibrium positions. After a few seconds a 5-7-5-7 defect cluster first appears, which dissociates by further agitation into two isolated dislocations [Figure S7(a)]. When the tracer is displaced even further one isolated dislocation ``ionizes''  into two isolated disclinations [Figure S7(b)]. These topological defects represent huge stresses inside the crystal lattice. In successive frames, the defect cluster at the head of the string performs a gliding motion down the string and eventually settles into a well-defined interstitial configuration [Figure S7(c)]. Intuitively, the particle that occupies such an off-lattice site requires a neighbouring particle to displace from its lattice site. The succession of such particle hopping events from their original lattice positions facilitates further cooperative motion. 

These experimental observations, substantiated with computer simulations, illustrate previously unexplored mechanisms of stress relaxation in soft crystalline solids. However, if these are true stress relaxation mechanisms, rather than mechanical instabilities caused by the driven particle, they should also exist in colloidal crystals excited by thermal fluctuations alone. Interestingly, our computer simulations indeed show that these unusual collective modes also emerge in crystals excited solely by thermal fluctuations. While such string-like motions of cooperatively moving particles have also been predicted to occur in amorphous systems such as supercooled liquids~\cite{donati1998stringlike}, their appearance in crystalline solids is remarkable. 

\begin{figure*}
\begin{center}
\includegraphics[width=0.9\textwidth]{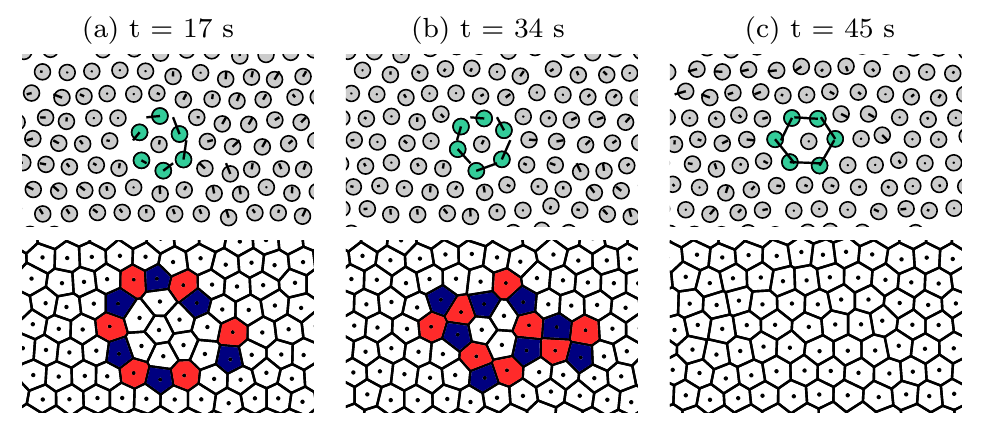}
\caption{Time lapse sequence of simulated configurations (top row) and corresponding Voronoi tesselations (bottom row) showing a spontaneous loop rearrangement (green particles) occurring without any driven perturbations. We have superimposed lines that connect the current particle positions with the initial defect-free configuration at $t = 0 s$.\label{fig:thermalloop}}
\end{center}
\end{figure*}

While a detailed investigation of these thermal collective modes in fragile crystals is subject for future research, we show one example of such a thermally-excited closed loop of  collectively moving particles in Figure \ref{fig:thermalloop}. A large thermal agitation causes the formation of a circular grain boundary in which the particles move in a cooperative ring-like fashion [Figure \ref{fig:thermalloop}(a)]. No difference is observed between the circular grain boundaries formed by means of mechanical excitations or those resulting from large thermal agitations. The formation of such circular strings of 5-7 defects is due to the small tilt in orientation between the bulk crystal lattice and the cooperatively moving particles. The rotational motion correlates with the effective annihilation of the 5-7 defects [Figure \ref{fig:thermalloop}(a-c)]. Similar as under mechanical perturbations, the particles become an integral part of the crystal again once the loop has closed [Figure \ref{fig:thermalloop}(c)]. Also the second type of rearrangement loops, which proceed through the formation, dissociation and relatively fast recombination of a vacancy-interstitial pair are observed in thermal equilibrium.

\begin{figure*}
\begin{center}
\includegraphics[width=1.0\textwidth]{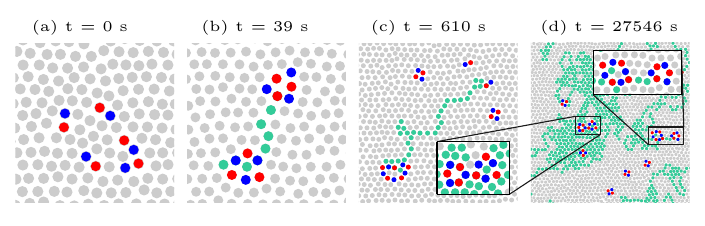}
\caption{Time lapse sequence of simulated configurations showing a spontaneous string of rearrangements (green particles) occurring without any driven perturbations. Particles with 5 or 7 nearest-neighbours are colored blue and red, respectively. \label{fig:thermalstring}}
\end{center}
\end{figure*}

Finally, the open-ended strings of collectively moving particles due to the spontaneous creation of dissociated vacancy-interstitial pairs are also found to exist in purely thermal systems, as shown in Figure \ref{fig:thermalstring}; here, we color-code particles with 5 or 7 nearest-neighbours blue and red, respectively. Particles that have irreversibly displaced one lattice spacing or more are color-coded with green. During the nucleation of the open-ended rearrangement string a region dense in defects forms inside the crystal [Figure \ref{fig:thermalstring}(a)]. As we do not apply any driving to the system, the formation of such a locally disordered zone is the direct effect of a large and local thermal fluctuation. The disordered region quickly settles into a clear vacancy and interstitial [Figure \ref{fig:thermalstring}(b)]. The time series clearly shows that the formation of a single self-interstitial has a profound effect on the single particle displacement through the diffusion of both the vacancy and the interstitial. After many particle rearrangements have occurred the vacancy and interstitial come into proximity again and annihilate in successive frames, as shown in the left zoom of Figure \ref{fig:thermalstring}(d). In the meantime, a new vacancy and interstitial have just nucleated somewhere else inside the crystal, as shown in the right zoom of Figure \ref{fig:thermalstring}(d).

\section{Discussion}
Our data show that large fluctuations inside a fragile crystalline solid are relaxed in the form of collective and cooperative motions of the constituent particles. For both large driven and thermal agitation we observe identical rich collective stress relaxation mechanisms, mainly in the form of open rearrangement strings through the motion of dissociated vacancy-interstitial pairs, and closed rearrangement loops through either rotational motions in circular grain boundaries or fast vacancy-interstitial pair dissociation-recombination reactions. Using optical tweezers we can excite these rare collective dynamics that are otherwise often kinetically restricted by large activation barriers.

The closed ring-like cooperative fluctuations we observe are a very efficient pathway to relieve both mechanical and thermal stresses [Figure \ref{fig:fig4}(c) and Figure \ref{fig:thermalloop}(c)]. Therefore, these modes provide additional stability to the crystal. For closed loop rearrangements, the highly localized and cooperative particle motions form an efficient pathway to annihilate defects and re-establish the crystalline structure. For the open string-like rearrangements, however, the local crystal lattice remains imperfect; topological disorder, in the form of vacancies and interstitials, is introduced into the lattice [Figure \ref{fig:stringgrowthexp} and Figure \ref{fig:thermalstring}]. Even though we observe closed loop rearrangements to be more abundant, both without and in the presence of driven perturbations, the open-ended strings have a much more profound influence on the single particle mobility. Whereas the closed loop rearrangements involve small groups of typically 3 to 16 particles [Figure S4 and S5], the open-ended rearrangement strings can cause many particles ($N \gg 500$) to exchange lattice sites irreversibly in a sequential manner. 

Despite the good agreement between experimental data and simulations, we find some subtle quantitative differences between experiments and simulations, especially for denser crystals [Figure \ref{fig:fig3}]. These differences might result from the break-down of the assumption of pair-wise additivity when the typical distances between particles become smaller~\cite{merrill2009many} or due to gradual changes in screening length over time due to leaching of ions from the glass sample chambers. 

In conclusion, we have shown, using a combination of experiments and simulations, how stresses relax in two-dimensional soft colloidal crystals, which are actively driven out-of-equilibrium. 
In conjunction with this instability we observe rich collective stress relaxation mechanisms, mainly in the form of open rearrangement strings through the motion of dissociated vacancy-interstitial pairs, and closed rearrangement loops through either rotational motions in circular grain boundaries or fast vacancy-interstitial pair dissociation-recombination reactions. Surprisingly, these unusual collective rearrangements are not restricted to crystals that are actively perturbed; computer simulations reveal that the same modes also exist in fragile crystals excited through thermal fluctuations alone. Our data illustrate how both large thermal and driven excitations in fragile crystals are relaxed through collective and activated modes, shedding new light on the origins of the stability of these fragile solid states.

\section{Methods}

\subsection{Colloidal crystals}
We use charged colloidal particles, consisting of poly(methyl methacrylate) with a surface layer of poly(hydroxystearic acid), synthesized following standard protocols~\cite{elsesser2010revisiting}, with a diameter $\sigma$ = 2.8 $\mu m$. When suspended in a 5 mM solution of sodium di-2-ethylhexylsulfosuccinate (AOT) in anhydrous dodecane, the particles acquire surface charges~\cite{hsu2005charge,sainis2008electrostatic}, resulting in long-ranged soft repulsive interactions. Using the approach described in Ref.~\cite{masri2011measuring} we find that the interactions are fitted well by a repulsive Yukawa potential with an inverse screening length of $\kappa=0.8$ $\mu m^{-1}$. Crystals are formed, at area fractions in excess of $\phi_A = 0.13$, by confining the particles in a quasi two-dimensional monolayer through gravity; we estimate a gravitational length $l_g \approx$ 80 nm, which is approximately 1$\%$ of the typical interparticle distance.

\subsection{Optical tweezing experiments}
Our optical tweezers consist of a 1.5W Nd:YAG laser, which generates a Gaussian diffraction-limited with $\lambda = $ 1064 nm. The laser is attenuated to 3 mW and guided into a set of acoustico-optic deflectors, which are used to steer the optical trap precisely in both $x$ and $y$ directions. The trapping beam enters the sample through through a 60x, high-NA, water-immersion objective. We optically trap a single colloidal particle within the crystal; local perturbations are induced by sinusoidally oscillating the particle, along one of the hexagonal crystal axes, around its equilibrium position with a frequency of $f$ = 0.005 Hz and amplitudes ranging from $A = 2.4$ $\mu m$ to $12.4$ $\mu m$, which equals between $1/3$ and $2$x the typical lattice spacing. With these parameters, the imposed motion of the trapped particle occurs on time scales similar to those of the measured Brownian time scale, thus minimizing hydrodynamic effects. Moreover, at the trapping wavelength of 1064 nm, dodecane is virtually transparent, which minimizes laser-induced heating of the sample. We systematically discard data in which profound out-of-plane motion occurs. Images, with a field of view of $144 \mu m \times 73 \mu m$, are obtained at 25-50 fps using bright field microscopy. Particle coordinates are determined using a standard tracking algorithm~\cite{gao2009accurate} for subsequent analysis. To avoid memory effects, we analyze only the first period of each experimental cycle; for subsequent measurements different locations within the extended crystal are used.

\subsection{Brownian dynamics simulations}
We perform Brownian dynamics simulations of $N = 2500$ charged particles in a box with aspect ratio $2 : \sqrt{3}$ and periodic boundary conditions. The colloid-colloid interaction is represented by a Yukuwa potential. We parameterise the potential using the method as described in Ref. ~\cite{masri2011measuring} and find good agreement with values reported in Ref.~\cite{hsu2005charge,roberts2008electrostatic}, yielding $\beta U_0=235$ and $\kappa \sigma =2.25$. The overdamped equation of motion for an undriven particle $i$ with position $r_i$ is given by $\eta r_i(t) = F_i(t)+R(t)$, where $R(t)$ is the random thermal force, and $F(t)$ is the total interaction force excerted on particle $i$ and the damping coefficient $\eta=1$. The driven particle $d$ is displaced sinusoidally ${\bf r}_{tracer}(t) = {\bf r}_{tracer,0}+A \sin(2 \pi f t)\hat{u}_{lattice}$ along its equilibrium lattice site $r_{tracer,0}$ in the direction of the lattice axis $\hat{u}_{lattice}$ with a fixed period $T=1/f=500000$ BD steps and an amplitude $A$. The hydrodynamic interactions and out-of-plane motions are neglected. We translate the simulation time to a physical time by comparing the time of self-diffusion in experiments ($t_{self}^{exp} \approx 11.2s$) and simulations ($t_{self}^{sim} \approx 20000$ BD steps).   
 
\section{Acknowledgments}
We thank H. van den Broek for help with the construction of the optical tweezers, and R. Wegh for developing software to operate the optical tweezers. J. Sprakel and J. van der Gucht thank the Netherlands Organisation for Scientific Research for financial support.

\end{document}


\title{Supporting information accompanying: \\
Highly cooperative stress relaxation in two-dimensional soft colloidal crystals}
\date{\today}
\author{B. van der Meer$^{1,2}$, W. Qi$^{2}$, R. Fokkink$^{1}$,\\ J. van der Gucht$^{1}$, M. Dijkstra$^{2}$ and J. Sprakel$^{1}$\\
\\$^1$Laboratory of Physical Chemistry and Colloid Science,\\
Wageningen University, Dreijenplein 6, 6703 HB Wageningen, The Netherlands
\\$^2$Soft Condensed Matter, Debye Institute for Nanomaterials Science,\\ 
Utrecht University, Princetonplein 5, 3584 CC Utrecht, The Netherlands\\}

\maketitle

\begin{sfigure*}[p]
\begin{center}
\includegraphics[width=0.5\textwidth]{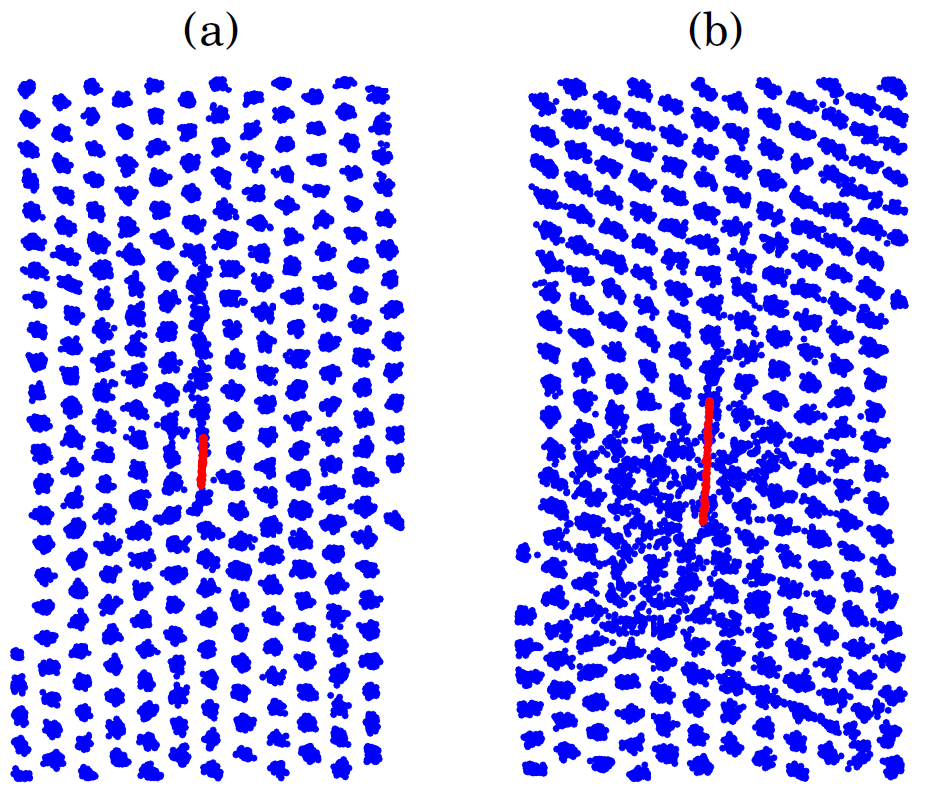}
\caption{\label{fig:respondsexp} Experimental particle trajectories (blue) at two different perturbation amplitudes $A$ (red) showing two distinct responses. (a) For small perturbations $A = 4.9$ $\mu m$ ($\gamma_L =0.5$) the crystal responds elastically, while (b) for higher perturbation amplitudes $A = 12.0$ $\mu m$ ($\gamma_L =1.2$) plastic deformation is observed. Lattice spacings: $a \approx 6.3$ $\mu m$.}
\end{center}
\end{sfigure*}

\begin{sfigure*}[!p]
\begin{center}
\includegraphics[width=0.5\textwidth]{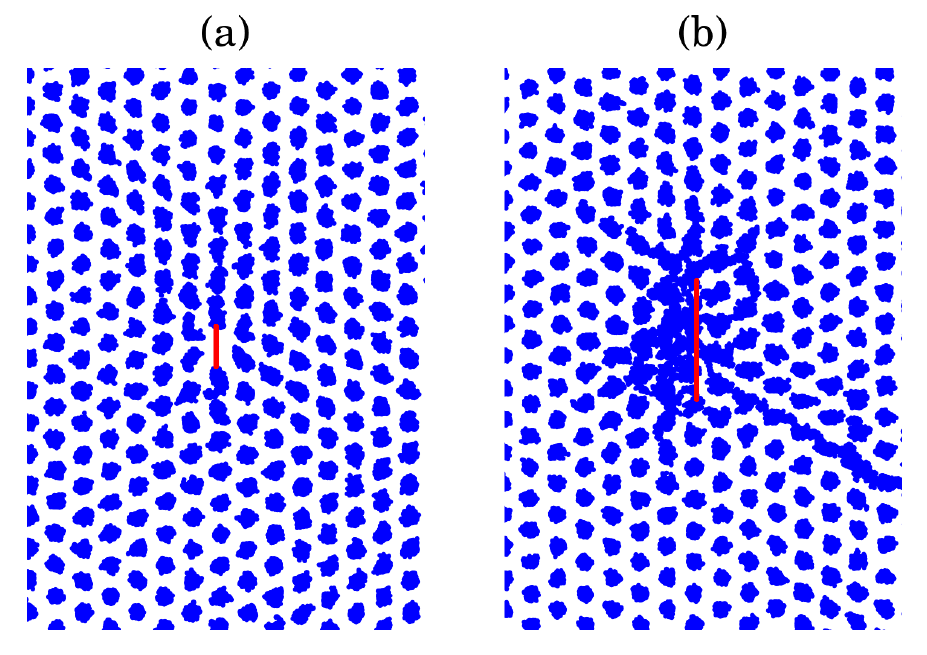}
\caption{\label{s:respondssim} Simulated particle trajectories (blue) at two different perturbation amplitudes $A$ (red) showing two distinct responses. (a) For small perturbations $A = 4.0$ $\mu m$ ($\gamma_L =0.4$) the crystal responds elastically, while (b) for higher perturbation amplitudes $A = 12.0$ $\mu m$ ($\gamma_L =1.2$) plastic deformation is observed. Lattice spacings: $a \approx 6.3$ $\mu m$.}
\end{center}
\end{sfigure*}

\begin{sfigure*}[!p]
\begin{center}
\includegraphics[width=0.7\textwidth]{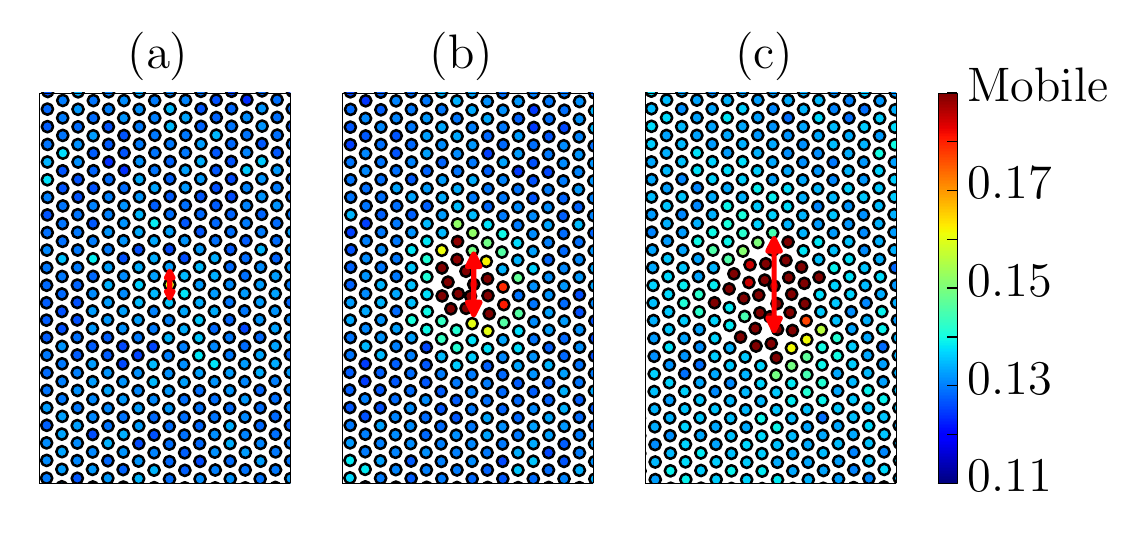}
\caption{\label{si:simmobile} Color map of the Lindemann parameter $L_m$ showing how the size of the localized mobile region depends on the perturbation amplitude $A$ in simulations. The particles are colored according to their value of the Lindemann parameter $L_m$ and color scale on the right. (a) For small amplitudes $A = 2.4$ $\mu m $ ($\gamma_L =0.25$) the crystal remains intact, while (b,c) for higher amplitudes $A = 7.2$ $\mu m$ ($\gamma_L =0.7$) and $ 12.0$ $\mu m $ ($\gamma_L =1.2$) a localized mobile region surrounds the driven colloid. Lattice spacings: $a \approx 6.3$ $\mu m$.}
\end{center}
\end{sfigure*}

\begin{sfigure*}[p]
\begin{center}
\includegraphics[width=0.6\textwidth]{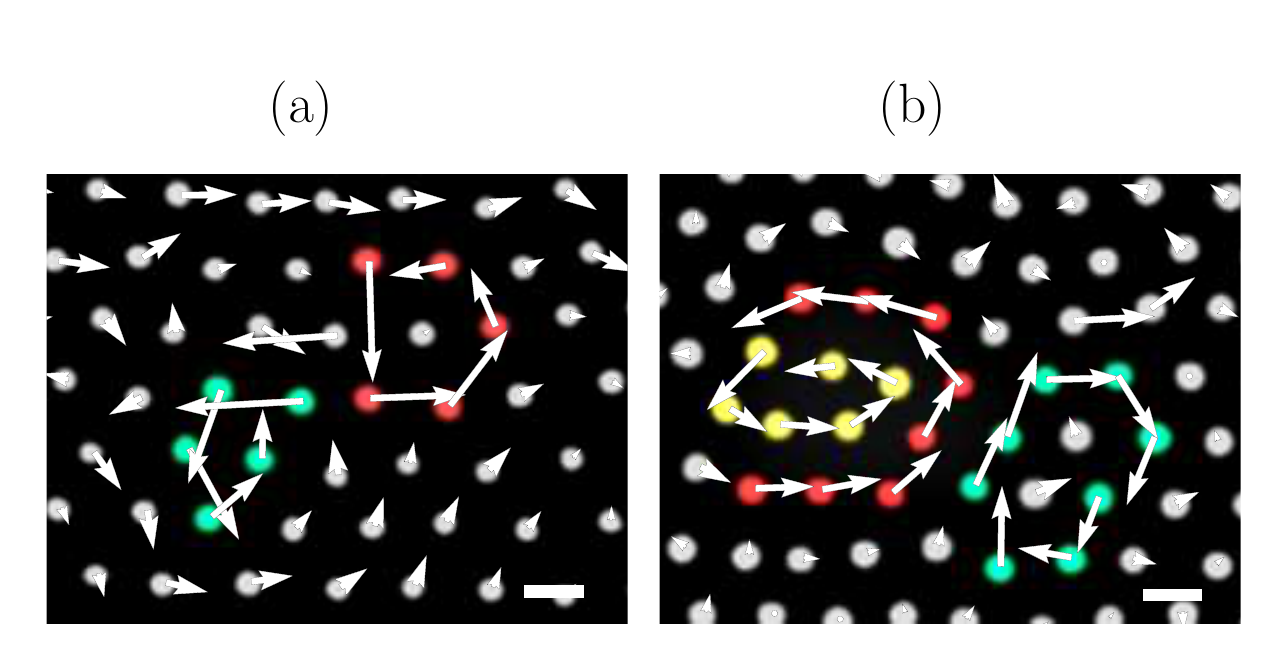}
\caption{\label{fig:fig5} Bright-field images of various types of string-like rearrangements. The superimposed quivers indicate the particle trajectories. (a) Two different types of five particle rearrangement loops. The red loop encircles a particle, whereas the green loop does not. (b) An eight (green) and six (yellow) particle rearrangement loop. The six particle loop is partially surrounded by an open string of rearrangements (red). Perturbation: $\gamma_L =1.2$. Scale bar: 5.0 $\mu m$.}
\end{center}
\end{sfigure*} 

\begin{sfigure*}[!p]
\begin{center}
\includegraphics[width=0.8\textwidth]{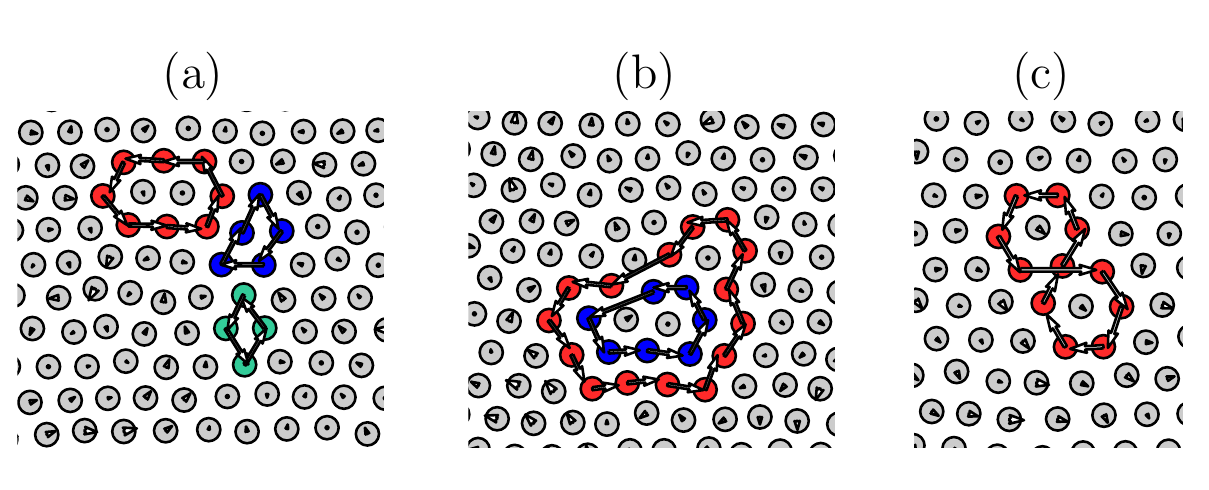}
\caption{\label{si:multiloopssim} Simulation snapshots of various types of string-like rearrangements. The superimposed quivers indicate the particle trajectories. (a) Frequently occurring rearrangement loops consisting of four, five and eight particles. (b) A seven particle rearrangement loop that is completely surrounded by a fifteen particle rearrangement loop. (c) An eleven particle rearrangement loop. Perturbation: $\gamma_L =1.2$.}
\end{center}
\end{sfigure*} 

\begin{sfigure*}[!p]
\begin{center}
\includegraphics[width=1\textwidth]{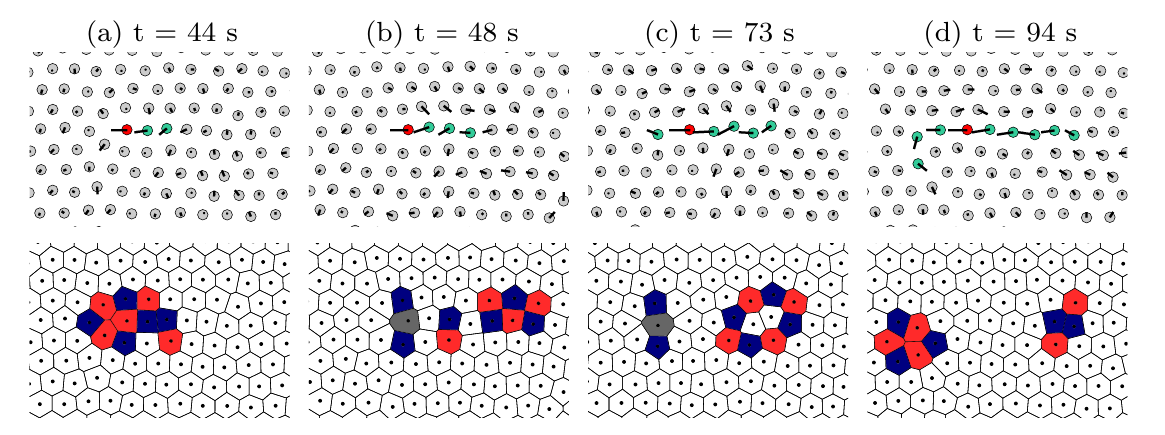}
\caption{Time lapse sequence of simulation snapshots and corresponding Voronoi tesselations showing a string of rearrangements, growing from both head and tail. When the trapped particle (red) is driven along one of the crystal axis, other, undriven particles start to move cooperatively as well (green). Perturbation: $\gamma_L =0.5$.}
\label{si:stringgrowthsim}
\end{center}
\end{sfigure*}

\begin{sfigure*}[!p]
\begin{center}
\includegraphics[width=0.65\textwidth,trim=0cm 0.2cm 0cm 1.2cm]{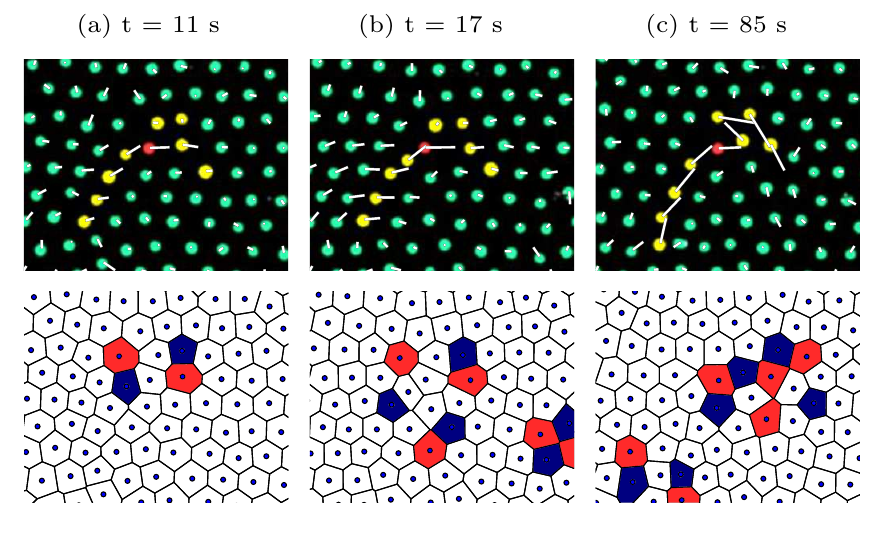}
\caption{\label{fig:fig6} Time lapse sequence of bright-field images and corresponding Voronoi tesselations showing an open-ended rearrangement string that grows from both head and tail. When the trapped particle (red) is driven along one of the crystal axis, other, undriven particles start to move cooperatively as well (yellow). In (a-c) we have superimposed lines that connect the current particle positions with the original equilibrium positions at $t = 0 s$. During the string-like cooperative motion many 5-fold (blue) and 7-fold (red) defects are formed. (a) In the initial stage of a developing string-like rearrangement, isolated dislocations (isolated 5-7 pairs) are formed. (b) One 5-7 pair dissociates into isolated disclinations. The defects at the head of the string perform a gliding motion along with the rearrangement string, driving further cooperative motion, and (c) eventually settle into a well-defined interstitial configuration. Perturbation: $\gamma_L =1.2$. Scale bar: 5.0 $\mu m$.}
\end{center}
\end{sfigure*}